# The Origin of (90) Antiope From Component-Resolved Near-Infrared Spectroscopy◊

F. Marchis[1,2,3], J. E. Enriquez[1], J.P. Emery[4], J. Berthier[3], P. Descamps[3], F. Vachier[3]

[1] SETI Institute, Carl Sagan Center, 189 Bernardo Avenue, Mountain View CA 94043, USA
[2] University of California at Berkeley, Department of Astronomy, 601 Campbell Hall, Berkeley CA 94720, USA
[3] Institut de Mécanique Céleste et de Calcul des Éphémérides, Observatoire de Paris, UMR8028 CNRS, 77 av. Denfert-Rochereau 75014 Paris, France
[4] University of Tennessee at Knoxville, 306 EPS Building, 1412 Circle Drive, Knoxville, TN 37996, USA

Pages: 46

Tables: 1

Figures: 11

*Corresponding author:*
Franck Marchis
Carl Sagan Center at the SETI Institute
189 Bernardo Avenue
Mountain View CA 94043, USA

fmarchis@seti.org
Phone: +1 650 810 0236
Fax:   +1 650 961 7099

---

◊ Based on observations from 382.C-0045 collected at the European Southern Observatory, Chile

**Abstract:**

The origin of the similary-sized binary asteroid (90) Antiope remains an unsolved puzzle. To constrain the origin of this unique double system, we recorded individual spectra of the components using SPIFFI, a near-infrared integral field spectrograph fed by SINFONI, an adaptive optics module available on VLT-UT4. Using our previously published orbital model, we requested telescope time when the separation of the components of (90) Antiope was larger than 0.087", to minimize the contamination between components, during the February 2009 opposition. Several multi-spectral data-cubes in J band (SNR=40) and H+K band (SNR=100) were recorded in three epochs and revealed the two components of (90) Antiope. After developing a specific photometric extraction method and running an error analysis by Monte-Carlo simulations, we successfully extracted reliable spectra of both components from 1.1 to 2.4 µm taken on the night of February 21, 2009. These spectra do not display any significant absorption features due to mafic mineral, ices, or organics, and their slopes are in agreement with both components being C- or Cb- type asteroids. Their constant flux ratio indicates that both components' surface reflectances are quite similar, with a 1-sigma variation of 7%. By comparison with 2MASS J, H, K color distribution of observed Themis family members, we conclude that both bodies were most likely formed at the same time and from the same material. The similarly-sized system could indeed be the result of the breakup of a rubble-pile proto-Antiope into two equal-sized bodies, but other scenarios of formation implying a common origin should also be considered.

1. **Introduction**

Among the ~190 recently discovered binary systems, (90) Antiope is certainly one of the most puzzling. Its large-amplitude and U-V-shaped lightcurve was reported by Hansen et al. (1997). In 2000, adaptive optics (AO) observations from the W.M. Keck II telescope, by Merline et al. (2000), revealed that this asteroid was composed of two similarly sized components. To date, about 30 similarly-sized binary systems (referred to as "double asteroids") have been detected in the main asteroid belt by their lightcurve signatures (Marchis et al. 2006). However, apart from (90) Antiope, none of them have been resolved using the high angular imaging capabilities of the Hubble Space Telescope (HST) or ground-based telescopes equipped with AO systems. The schematic diagram shown in Figure 1 describes several well-characterized double systems located in the main asteroid belt. It is notable that (90) Antiope, being the only large double binary currently known, is unique among this small sub-population of binary asteroids.

[INSERT FIG. 1]

Combining lightcurves and adaptive optics observations collected with 8-10m class telescopes, Descamps et al. (2007) determined the components' orbital parameters and an overall shape solution of two tidally-locked ellipsoidal components with an average diameter of ~90 km. They derived a low bulk density of 1.25 g/cm$^3$, which they interpreted as being due to a rubble-pile interior with a macro-porosity of 30%. An additional study published more recently by the same group (Descamps et al. 2009), but using additional data taken during mutual events in 2007-2008, confirmed the accuracy of the orbital parameters and revealed the presence of a 68-km diameter crater on one of the components. This geological feature could be the aftermath of a collision against a 100-km, porous proto-Antiope that split it into two components by spin up.

(90) Antiope is a C-type asteroid member of the Themis family, a collisional family identified by Marzari et al. (1995). This group of C- and B-type members is located in the outer main belt and is believed to be the result of the catastrophic disruption of a large 450-km diameter parent body 2.5 +/-1 Byrs ago (see the recent work by Nesvorny et al. 2005). After the recent detection of water ice on the surface of 24 Themis (Campins et al. 2010 and Rivkin and Emery, 2010), Castillo-Rogez and Schmidt (2010) discussed the possibility of the large proto-Themis being a differentiated body made of a mixture of ice and rock. In this case, (90) Antiope could be an "icy asteroid," with a significant portion of water ice in its interior explaining its low bulk density and the equilibrium shape of its components. This idea is supported by a spectroscopic study of 36 members of the Themis family (Florczak et al. 1999) which reveals a large range of C-complex sub-classes (e.g. C- B- G- F- types) and the presence of different degrees of aqueous alterations. The Themis family members could have been formed from an already differentiated body.

Polishook et al. (2009) conducted an extensive visible photometric and spectroscopic study of (90) Antiope in January 2008 while the system was seen nearly edge-on from Earth, presenting total eclipse configurations. All spectra taken before and after two eclipses separated by 1.5 rotations are featureless and their spectral slopes are stable within 3%. Descamps et al. (2009) fitted visible lightcurves of (90) Antiope collected since 1996, from a solar phase range from -15 deg to +15 deg, with a similar set of Hapke photometric parameters. These works, based on visible photometric and spectroscopic data, suggest that the two components could be similar in composition and underwent the same physical processing, implying a common origin.

To better constrain the origin of the system and determine whether or not the components are identical in composition and age, we conducted a comparative near-infrared spectroscopic study of the two components of (90) Antiope. Using the high angular resolution provided by adaptive optics systems on 8-10m class telescopes, combined with the spectroscopic capabilities of integral field unit, we are now able to simultaneously record the spectra of both components. The first such component-resolved spectra of a binary asteroid was recently presented by Laver et al. (2009), revealing that (22) Kalliope and its satellite, Linus, are remarkably similar in near-infrared surface reflectance.

Figure 2 contains the visible and near-infrared spectra of several C-complex asteroids, and X- and S-type asteroids from DeMeo et al. (2009). Near-infrared wavelengths are more appropriate for revealing differences in composition between two components of a binary asteroid since i) the wavelength range is broader, facilitating determination of the spectral slope, ii) the near-infrared range displays two absorption bands centered at ~0.9 µm and 2.1 µm, due to mafic materials on the surface.

[INSERT FIGURE 2]

In this work, we report the complete analysis of component-resolved spectroscopic observations of (90) Antiope. In Section 2, we describe our observational strategy, the data reduction, and the quality of the final data. Section 3 contains a description of our photometric extraction method, and the Monte-Carlo simulations that we performed to extract the flux and to better estimate the final error in the collected data. The final spectra and their comparison and the implication of the origin are discussed in Section 4. We conclude in Section 5 by contrasting this work with other spectroscopic studies of binary asteroid systems in the main-belt.

## 2. Observations

### 2.1 Strategy of observation

(90) Antiope is unique among similarly sized binary asteroids since it is the only one which can be resolved with a 8-10m class telescope equipped with adaptive optics. Both components of the system are orbiting around their center of mass, describing a circular orbit with a separation of 171 km in 16.5051 h (Descamps et al. 2007). The binary system is orbiting around the sun with a semi-major axis of 3.156 AU and describing an eccentric (e=0.156) and slightly inclined (i=2.22 deg) orbit. The maximum angular separation between the two components observed from Earth varies from 0.090 to 0.144 arcsec.

In 2008, the (90) Antiope system was predicted to be at opposition on March 2, 2009 with a distance from Earth of 2.594 AU, corresponding to an angular separation of 0.090 arcsec between both components. The UT4 of the Very Large Telescope is equipped with SINFONI, an adaptive optics system, and an integral field spectrograph named SPIFFI (SPectrometer for Infrared Faint Field Imaging). SPIFFI is an advanced reflective image slicer and an efficient spectrograph equipped with a 2K x 2K Rockwell Hawaii infrared array 2RD detector, four gratings (J, H, K, and H+K) providing a spectral resolution between 1,500 and 4,000 (Eisenhauer et al. 2003), and three fields of view of 0.8 arcsec, 3.2 arcsec and 8 arcsec. The Estimated Time Calculator (ETC) indicated that the SINFONI AO could provide a stable, good correction (Strehl Ratio~35%) on a bright target like (90) Antiope (whose predicted magnitude in visible was 13.3), reaching an angular resolution of ~0.050 arcsec at 1.6 µm. Because (90) Antiope could be observed for 5 hours from ESO-Paranal with an airmass lower than 1.6, we requested telescope time to study the components of Antiope by comparative near-infrared spectroscopy during the 2009 opposition.

We predicted the appearance of the system using the physical ephemeris developed by the IMCCE and the orbital solution published in Descamps et al. (2009). We used the Descamps et al. (2009) model since it provides a refined pole solution $\lambda_0$= 199.5 ± 0.5 deg and $\beta_0$= 39.5 ± 5 deg in J2000 ecliptic. Since the system was in a nearly edge-on configuration, we determined several optimal observing windows, requiring that the two components' separation was larger than 0.087 arcsec (to minimize the effect of contamination between components) and that the system was observable from the VLT with an airmass less than 1.6 (to provide a good AO correction). Eleven windows of opportunity lasting 2.75 h, including 6 high-priority windows close to (90) Antiope's opposition, were listed in the proposal. We requested two epochs of observations to be conducted in service mode. Each run was defined in a similar manner. The smallest plate scale of the SPIFFI instrument, with a field of view of 0.8 x 0.8 arcsec and a spatial scale of 12.5 x 25 mas, would be used. The observations would consist of two consecutive spectra taken through the J grating (R=2,000) and the H+K grating (R=1,500) with a total exposure time and overhead of 2.7 hours.

The OPC of the European Southern Observatory notified us in July 2008 that the proposal (numbered 382.C-0045) was accepted and that 5.4 hours of observation ranked in category A was allocated to us. We added observations of a nearby solar analog star named BD+11 2282 (G0 type with RA= 10h 49min 02.98s DEC=+10° 25' 26.9" in ICRS J2000, B=10.45, V=9.94, K=8.50), for photometric calibration, airmass correction, and to estimate the quality on the data. Total exposure time per grating on the star was set to 240 s. As a consequence, the total exposure time on the target was reduced to 600 s per grating per epoch, reaching a total requested time of 5.4 hours, including telescope and instrument overheads. To take full advantage of the angular resolution provided by the AO system we also requested that the observations be taken at an airmass lower than 1.6 and under exterior seeing conditions better than 1.0 arcsec.

**2.2 Observations and basic data-processing**

Table 1 summarizes the data collected under our observing program. Three epochs of observations were collected on Feb 1, Feb 3 and Feb 21, 2009. Each epoch is composed of two observing blocks of the solar analog star BD+11-2282 with the J and H+K gratings, followed by observing blocks of (90) Antiope in J and H+K, and concluded by two more observing blocks of the solar analog.

[INSERT TABLE 1]

The data were processed using the SINFONI/SPIFFI data reduction pipeline developed by the ESO team. Details about the data reduction can be found in the SINFONI data reduction cookbook[1] and the SINFONI user manual[2]. The methods used by the pipeline to generate calibrations are relatively straightforward:

- Each night, a bad pixel map is created by analyzing the deviation of a pixel through a stack of frames.

- A master dark is generated from input frames with the same integration time.

- A master flat field is computed by subtracting frames taken with the dome light off from frames taken with dome light on. Bad pixels are then interpolated and the master flat is normalized.

- The detector distortion and the positions of the 32 continuum slit spectra are determined by subtracting 79 calibration frames illuminated by 75 fibers. This part of the calibration is routinely measured once per month and after each instrument intervention.

- Arc frames are created to derive the wavelength solution and calibrate the observations. They come in stacks of several pairs of lamp-on and lamp-off frames. They are routinely measured every morning. The product tables of these calibrations are quality-checked on the mountain and at the ESO quality control office in Germany.

---

[1] http://www.eso.org/sci/facilities/paranal/instruments/sinfoni/doc/VLT-MAN-ESO-14700-4037.pdf
[2] http://www.eso.org/sci/facilities/paranal/instruments/sinfoni/doc/VLT-MAN-ESO-14700-3517_v87.pdf

The science products consist of a wavelength-calibrated image data cube built by subtracting object-sky frames recorded in random jittering. For each object frame the sky frame closest in time is selected and subtracted from the object. After removing bad pixels by interpolation and correcting the pixel-to-pixel response by dividing with the flat-field, each slitlet spectra is corrected for distortion. Using the known positions of the slitlet spectra, the ESO pipeline builds a cube composed of imaging frames, with a pixel scale of 12.5 mas, and the wavelength range in the Z-direction. The H+K and J grating cubes are composed of 2172 frames and 2236 frames, respectively. The resulting cubes are weighted by integration time and averaged together. The final cubes have variable spatial dimensions defined by the angular size and extension of the targets. For (90) Antiope observations, the final frames are ~100 x 100 pixels (so 0.125 x 0.125 arcsec). For the solar analog observations, the cube is smaller in the spatial direction with a dimension of ~60 x ~60 pixels.

## 2.3 Data quality

Figure 3 shows one slice extracted from the H+K cube (at 1.55 µm) and one slice of the J cube (at 1.13 µm) for each epoch of observations. A visual inspection of these frames confirms that there is an excellent agreement between the observations and the Descamps et al. (2009) orbital model shown in the last column.

**[INSERT HERE Fig. 3]**

The first observation recorded on February 1, 2009 was taken inside a selected window of observations but with poor seeing conditions (from 1.0 to 1.3 arcsec). The angular resolution indicated in Table 1 is measured by estimating the full width at half-maximum (FWHM) of the larger component of (90) Antiope or the solar analog star at 1.13 µm for the J grating and 1.55 µm for the H+K grating. The solar analog is located less than 4 degrees from (90) Antiope, so its FWHM should reflect the angular resolution and its variability at the time of the (90) Antiope observation. It varies from 0.063 arcsec to 0.215 arcsec on the star and is estimated to be 0.076 arcsec on (90) Antiope for the H+K grating.

The second epoch, taken two days later on February 3, was not listed as a window of opportunity in our program. The angular separation between the two components of (90) Antiope is estimated to be ~0.025 arcsec for the H+K grating. Considering the angular resolution measured on the solar analog star (from 0.071 to 0.108 arcsec) these observations are not usable since the system is not resolved.

Only the third epoch on February 21 was optimal under both criteria, with an estimated separation larger than 0.080 arcsec and excellent seeing conditions estimated to less than 0.7 arcsec. The angular resolution estimated on the larger component of (90) Antiope and on the star is quite stable, varying from 0.060 to 0.073 arcsec for the H+K grating and 0.058 to 0.071 arcsec for the J grating. Additionally, we should expect higher quality for these data since the observations were conducted when (90) Antiope was closer to its opposition.

The goal of this work is to compare the spectrum of each component of the system over a large wavelength range in the near-infrared. Because of their poor quality, we discarded the data taken on Feb 03 with the J and H+K gratings and on Feb 01 with the J grating. It should be remembered that the images (slices in the data cube) are built from a large series of slitlet spectra in a 2K x 2K frame after numerical processing. An inspection of the reconstructed image of the star shows that the maximum peak value varies significantly, since the profile of the centroid is not symmetric. Consequently, we cannot estimate the flux ratio between the two components simply by comparing their peak values. The quality of the data depends on seeing conditions and varies within each spectral cube.

In Fig. 4 we display several extracted slices of the H+K cubes from Feb 21 and Feb 01, displayed with the same color cut. In the Feb 21 observation, angular resolution seems to increase with wavelength from 1.44 to 2.11 µm, as expected under good seeing conditions, since the theoretical angular resolution of a telescope varies as $\lambda/D$. The frame extracted at 1.93 µm has a poor SNR since it is taken in the middle of an atmospheric absorption window separating the H and K bands. In the middle column of Fig. 5 we plotted the characteristics of this cube. The SNR reaches a peak value of 140 at 1.5 µm and decreases towards larger wavelengths due to the combined effects of lower angular resolution and higher background noise. The Feb 01 data cube is poorer in quality, with a peak SNR of 40 at 2.1 µm, similar to the SNR of the J band observations taken on Feb. 21.

[INSERT HERE Fig. 4]

We calculated two types of FWHM values for the (90) Antiope observations:
- The radial FWHM (along the axis connecting the two components)
- The tangential FWHM (perpendicular to the radial axis)

The behavior of the radial and the tangential FWHM differ between epochs of observation and wavelength ranges. On Feb 21, each H+K radial FWHM increases with wavelength, and the ratio of both FWHMs is close to 1, implying that the AO correction is better and more stable at longer wavelengths. The Feb 21 J band observations are of poorer quality since the PSF are asymmetrical (FWHM(Radial)/FWHM(Tangential) ~ 0.8) and have a FWHM(Radial) of ~6 pixels. This is somewhat expected since AO systems provide poorer correction at shorter wavelengths. The Feb 01 data in the H+K grating are quite similar to the J band data taken on Feb 21 (SNR~40, FWHM~6 pixels~75 milli-arcsec), implying that the Feb 01 data were taken under poor seeing conditions.

[INSERT here Fig. 5]

3. Data Analysis

3.1 Photometric extraction

Figure 5 clearly illustrates the challenge of our analysis. Our goal is to be able to independently extract an accurate estimate of the flux per wavelength for each component, and via comparison to constrain their surface compositions. The components are very close to each other with an angular separation of 0.080 arcsec, corresponding to less than 1.5 elements of resolution. Since the data quality varies by night, by filter, and along each cube of observations due to both atmospheric transparencies and the instrument PSF, we need to independently extract the flux from each frame using a numerical fitting process specially developed for this kind of data.

Our algorithm, written exclusively using IDL, consists of fitting each component's centroid by two Gaussian functions to mimic the profile of an AO system. An AO PSF is composed of a coherent peak, with a FWHM close to the diffraction limit of the telescope, surrounded by an extended halo due to the imperfect correction of the AO system. Our final fit is a combination of a wide Gaussian function with a FWHM of ~0.075 arcsec and a narrow Gaussian function with a FWHM of ~0.045 arcsec, which are determined by fitting pixels with an intensity larger than 60% (for the narrow Gaussian function) and 25% (for the wide Gaussian function) of the peak intensity. The final fit is obtained by multiplying these two Gaussian functions by two empirical coefficients (0.14 ± 0.01 for the narrow Gaussian and 0.7 ± 0.1 for the wide Gaussian), which are estimated for each night and each filter.

An example of the fitting process for the frame at 1.22 μm taken on Feb 21 is shown in Fig. 6. This is clearly not the best frame, with SNR~40, FWHM(radial)~6.4 and 6 pixels (for components A and B), and FWHM(tangential)~8 pixels (from the first column of Fig. 5). Figure 6 shows the 1.22 μm image and a fit (dotted line) along the radial direction of the system. The centroid of each component is independently fitted on this frame by two Gaussian functions, shown as a solid line in the figure and the inset picture. The residual is shown along the radial fit and on the image (leftmost bottom panel in Fig. 6). The average value of the residual is ~7% of the peak intensity and remains uniform around each fitted function, indicating that the relative intensity of each component should be well recovered.

[INSERT here Fig. 6]

This fitting process was applied to 2172 frames of the H+K cube recorded on Feb 01 and Feb 21 and to the 2236 frames composing the J cube recorded on Feb 21. Before executing the time-costly (4-8h) and automated fit, each frame of a cube was checked for scientific usability. The following frames were rejected:

- The first ~100 and last ~100 and 400 frames of the cubes in H+K and J gratings respectively since they have a low SNR
- ~300 frames located between the H and K band filters, because of their low SNR
- Those contaminated by cosmic rays, or with too many bad pixels and artifacts
- Those with a SNR lower than 3

On average, 25% and 22% of the frames in the H+K and J cubes, respectively, were rejected based on these criteria.

Variations in the refractive index of Earth's atmosphere in the near infrared cause the peak positions of the components to vary between frames. To speed up the fitting process, we first guessed the peak positions by interpolating the fitted positions of three averages of 50 consecutive frames (with a good SNR), taken at the beginning, middle and end of each cube. The dual Gaussian fit is then applied to each component of the (90) Antiope system. To derive the photometric flux, minimizing the remaining effect of contamination on the frames, we calculated the integrated photometric value by using a modified aperture photometry algorithm. The flux is estimated by averaging the flux values determined with an aperture of 1/3 the separation of the components and with an aperture of 2.2 × FWHM(radial) of the components. We tested this adjustment on simulated images and it seems to give the best estimate of the absolute flux encompassing the separation of the components and the image quality.

**3.2 Reflectance spectra**

After the extraction of the photometric measurements of each component, we derived spectra using standard procedures. Arc lamp observations, taken in the morning shortly after our observations and processed by the SINFONI team, were used to derive the wavelength of each frame in a cube. For each grating cube, we processed two spectroscopic observations of the solar analog BD+11-2282, taken before and after the observation of (90) Antiope and taken in a similar mode (see Table 1). We corrected both spectra for the refraction (90) Antiope by interpolating the spectra of BD+11-2282 recorded before and after the target to the same airmass. We derived the reflectance spectra by dividing the (90) Antiope spectra by their respective solar analogs.

Figure 7 shows the spectra of each component with J and H+K gratings on Feb 21 and with H+K on Feb 01, after resampling to a spectral resolution R of 150 in H+K and 200 in J. Due to the removal of poor-SNR frames on the edges of the cubes and between the H and K atmosphere windows, the final spectra are from 1.1 to 2.4 µm with two gaps from 1.35 to 1.45 and 1.8 to 1.95 µm. For comparison, we included an integrated spectrum taken from 1.0 to 2.5 µm using the IRTF/SPEX instrument and superimposed the average of both components' spectra in J and H+K gratings. The IRTF spectrum recorded on Jul 22 2006 from 13:43 to 14:45 UT was processed as described in Emery et al. (2005).

A visual inspection shows that the Feb 21 spectra of both components look identical and that the averaged spectrum is similar to the Jul 2006 data, validating the robustness of our photometric extraction process. Neither of the Feb 21 spectra display any absorption features of mafic minerals (proxene, olivine), ices ($H_2O$, $NH_3$, $CH_4$, CO) or organic materials. Such a featureless and relatively flat spectrum is the main characteristic of C-type asteroids in this wavelength range.

**[INSERT HERE FIG 7]**

**3.3 Photometric error analysis**

To estimate the error of our photometric extraction we performed a complex Monte Carlo simulation. We generated several synthetic images of (90) Antiope composed of 2 components with a flux ratio of 0.90 (expected from a relative size ratio of 0.94 from Descamps et al. 2009) and a center of light made of two Gaussian functions. The set of synthetic image is made of two component located at 60 degrees counter clockwise with a SNR of 25, 40, 100 and 160 and a "*proximity*" values from 0.8 to 1.2.

The "*proximity*" is defined as the ratio of the separation of the components with the FWHM of the centroid of one component. Because the measured FWHM of our data shown in row 2 of Fig. 5, is nearly identical for each component in one frame, we generated artificial images with the same characteristics. A *proximity* of ~1 corresponds to an angular separation between the two components equal to the angular resolution of the instrument; in other words, the system is fully resolved. Figure 8 shows several synthetic images generated with a SNR of 25 and 60 and a *proximity* ranging from 0.77 to 1.20 assuming that the noise is a combination of Gaussian noise and Poisson noise. They are indeed visually identical to the images extracted from the observations.

**[INSERT here Fig. 8 & 9]**

For each image extracted from our cubes of observations we measured the value of the SNR and the *proximity*. We generated 100 synthetic frames as shown in Fig. 8 after adding the noises. For each simulated frame, we ran our extraction algorithm as presented in Section 2.4 on and derived the photometric accuracy. Figure 9 summarizes the result of this Monte-Carlo analysis. The 1-$\sigma$ photometric accuracy varies from 3 to 9%. As expected, it is worst for a small *proximity* value and to a lesser extend for low SNR. With low *proximity* the fitting process is less accurate and the contamination effect is larger. It improves up to 3% for a *proximity* of 1.2 and a SNR larger than 40. The average value and the range of *proximity* and SNR that we measured on each data cube are labeled on Fig. 9. The Feb 21 H+K grating dataset is the best, with a SNR from 70 to 120 and a *proximity* from 1.03 to 1.08. The Feb 01 H+K data cube was the poorest, with a SNR of ~40 and a *proximity* ranging from 0.72 and 0.92.

By combining the results of our photometric extraction and the derived characteristics of each frame (SNR and *proximity*) with the result of our Monte-Carlo noise estimate analysis, we are now able to derive a formal error on the photometric extraction. We defined it by summing quadratically a systematic error defined by the difference in the ratio flux of measured on the components and the 1-sigma variance of this ratio for the set of 100 artificial images. The Feb 21 J and H+K spectra of (90) Antiope's components and the formal error are drawn in Fig. 10. The reflectance spectra of the components are quite similar to the average spectra of C- and Cb- type asteroids from the DeMeo et al. (2009) classification. Other C-complex classes such as Cg, Cgh, and Ch have a slope radically different to the spectra of (90) Antiope Components. The H+K spectrum taken on Feb 01, and shown in Figure 7, is obviously more problematic, with a slope outside the expected range for C-complex asteroids and with significant differences from the July 2006 integrated spectrum. Figure 3 indicates that both epoch observations were done with an almost identical configuration (with identical SEP latitudes and a difference in SEP longitude of ~20 deg). Our error analysis shows that data collected during this night suffer from low proximity (<0.8) due to the poor seeing conditions (Table 1 and Figure 8). In conclusion, it is certain that this reddening of the spectrum is not real but is due to technical limits in the photometric extraction for these poor resolution data.

**[INSERT FIG 10]**

## 4. Results and Discussion

**4.1 Spectra ratio**

The spectra of both components of (90) Antiope were successfully extracted using our algorithm on data from Feb 21 in both J and H+K grating modes. Their spectra are featureless and have a slope compatible with a C- or Cb- type composition. The ratio of the two components' spectra and the error estimates for both grating modes is plotted in Figure 11. This figure reveals a relatively constant ratio with a best-fit flux ratio of 0.87 and a 1-sigma variation of 0.07. The Roche ellipsoid model proposed by Descamps et al. (2009) has a size ratio of 0.94±0.03 corresponding to a projected surface ratio of 0.90, in agreement with our flux ratio measurement. Considering the relative uncertainty of 5% in the spectra ratio, it is difficult to extract more accurate information about the albedo of these components in the near-infrared. No difference in slope and no broad absorption bands can be seen on the ratio of these spectra, implying that the two components are indeed spectroscopically identical (to within 7%) in the near-infrared from 1.1 to 2.4 µm. The small periodic variations with a period of 0.04 µm and an amplitude of 0.06-0.07 seen mostly in the H+K band mode are most likely artifacts due to the instrument or confusion in the photometric extraction algorithm. The long-period (0.4 µm) variation in J band with an amplitude of 0.09 is due to the low quality of the AO data at short wavelength producing confusion in the photometric measurement.

**4.2 Discussion of the nature of the component surface**

The averaged spectrum of both components taken from 1.1 to 2.4 µm is plotted in Fig. 7 and confirms that (90) Antiope is possibly a C- or Cb- type asteroid. The near-infrared spectrum, which has a slope of ~0.17 unit of reflectance(UR)/µm, is redder than the average spectra of Cg- (0.12 UR/µm), Cgh- and Ch- (0.07 UR/µm) class asteroids from DeMeo et al. (2009) taxonomic class.

If (90) Antiope is indeed an icy remnant of the collision of a proto-Themis body as suggested by Castillo-Rogez and Schmidt (2010), based on its low bulk density of 1.28 g/cm$^3$ (Descamps et al. 2007), and if a subsurface ice reservoir supplied enough water to the surface, then the spectra of the components could plausibly display signatures of water ice with broad absorption bands centered at 1.25, 1.52, and 2.02 µm (e.g., Emery et al. 2005). The absence of those features in the present data places weak limits on the abundance of ice. No data have been published of (90) Antiope at longer wavelengths ($\lambda$ > 2.5 µm), so it is not possible to say whether or not it exhibits a 3.1-µm water ice band like 24 Themis (Campins et al. 2010, Rivkin and Emery 2010), which is also featureless in the 1.1 to 2.4 µm range.

It is interesting to discuss if identical reflectance automatically implies that the components' surfaces are identical in composition, were formed at the same time, and evolved in the same environment. The origin of (90) Antiope remains a mystery and at the present no computer simulations have successfully explained the origin of this unique large double binary asteroid. Descamps et al (2009) extrapolated from the detection of a large bowl-shape crater on one of the components that the system could have formed by fission. In this case, and because the proto-Antiope was a rubble-pile asteroid resulting from the disruption of Themis, the interior of both components should be quite homogenous and nearly identical. The binary system has evolved for 3.5 Gyr in the same environment in the outer part of the asteroid main belt, implying that (90) Antiope could be one of the most primitive rubble-pile bodies in our solar system. Spatially-resolved near-infrared spectroscopic and photometric studies of C-type asteroids are rare. Phobos, a satellite of Mars and historically classified as a C-type asteroid from visible photometric observations collected with the Viking spacecraft (Pang et al. 1978), has been studied in the near-infrared with multispectral sensors on board *Phobos 2* (Murchie et al. 1996) and more recently with CRISM aboard the Mars Reconnaissance Orbiter (Murchie et al. 2008). Both studies revealed a heterogeneous surface with blue and red terrains with measured slope variations of ~0.15 UR/μm and ~0.46 UR/μm, respectively, in a limited number of spectra (Murchie et al. 2008). As pointed out by Murchie and Erard (1996) the blue unit on the surface of Phobos has a similar slope and a low albedo as C-type asteroids. Figure 10 from Murchie and Erard (1996) shows visible and NIR spectra of selected regions on Phobos. It is quite interesting to see that the relative slopes of the red regions are constant to within less than 15% from 1.1 to 2.4 μm, implying that the variation in reflectance of the surface material of Phobos is small. The ratio of the spectra of the components of (90) Antiope derived from our study is shown to be uniform to within 7%. We can therefore conclude that if Phobos is indeed an analog to a C-type asteroid, then the small discrepancies between the near infrared spectra of the components of (90) Antiope (in Figures 9 and 10) are not in contradiction with the system having formed from the same material.

We could also wonder if this similarity in the near-infrared would be explained if both components are in fact captured fragments from the Themis family. Because our search for near-infrared spectra of Themis members showed that a handful of NIR spectra were recorded, we cannot perform a reliable comparison of the spectra of these family members with the Antiope components. As an alternative, we extracted from the 2MASS Second Incremental Release catalog 93 asteroids identified by Zappala et al. (1995) as Themis family members. We ran a Monte-Carlo simulation to identify the probability that these asteroids have the same color as the Antiope components. We computed the color distribution of these asteroids considering the 1-sigma error derived from the 2MASS color measurements in J, H and K bands and determined how many of them could have the same colors as Antiope A (J-K = -0.161, H-K = -0.204) and Antiope B (J-K = -0.119, H-K = -0.071) which were derived from our data. For a sample of 1,000 tries, we derived a probability that the two components of Antiope were captured from the Themis family of only 7% ± 2. This probability should be considered an approximation, and certainly an upper limit, since a comparison of the spectra from 1.1 to 2.4 mm should be more restrictive than a color comparison.

We can therefore conclude that it is very likely that these components formed from the same material and at the same time. It is, however, beyond the scope of our results to determine if the system's formation is due to an oblique impact, tidal splitting after a close encounter with a large asteroid, or any other scenario implying a common parent body.

### 5. Conclusion

We present in this work the first successful observations of (90) Antiope using SINFONI, an integral field unit available on one of the 8m-VLT telescopes equipped with adaptive optics at ESO Paranal. Even though the observations were recorded in three epochs near opposition from Feb 01 to Feb 21 2009, the angular separation of both components (0.090 arcsec) was very close to this instrument's diffraction limit (0.067 arcsec). We developed a specific photometric extraction algorithm to compare the flux of the components in the J and H+K grating modes (from 1.1 to 2.4 µm with two gaps from 1.35 to 1.45 and 1.8 to 1.95 µm). A formal error was found using Monte-Carlo simulations mimicking the images provided by the instrument. Only one epoch of observation, Feb 21, was considered good enough to provide a reliable estimate of the individual spectra of (90) Antiope's components.

The final, low-resolution (R~150-200) spectra of each component of (90) Antiope, shown in Fig. 7, indicate that they have a near-infrared reflectance typical of the C-type or Cb-type asteroid spectrum, characterized by the absence of features and a relative flatness (DeMeo et al. 2009). The average flux ratio of the spectra, plotted in Fig. 8 and estimated to be 0.87, is compatible with the components' relative size ratio as estimated by Descamps et al. (2007). The components' spectra are remarkably similar between 1.1. and 2.4 µm (identical within a 7% margin) and do not show any absorption features that could be due to the presence of mafic minerals, ices, or organics on the surface. Polishook et al. (2009) recently reported a visible-wavelength spectroscopic study of the system in a mutual event configuration, which similarly detected no variations of reflectance in their wavelength range. The addition of the near-infrared range, which includes absorption bands of materials detected on the surface of asteroids and shows a larger gradient of slopes among the C-group asteroid population (Fig. 2), is critical to our conclusion that it is very likely that the two components of (90) Antiope formed at the same time from the same material, as suggested by Descamps et al. (2009).

This work is the second published component-resolved spectroscopic study of a multiple asteroid, but is the first to study a similarly sized binary asteroid and a low-albedo C-type asteroid. Laver et al. (2009) used OSIRIS integral field Unit to compare the near-infrared reflectance spectra of the well characterized, large (D=166 km) main-belt asteroid (22) Kalliope and its 28-km satellite, Linus. (22) Kalliope, an M-type asteroid with a featureless reflectance spectrum in the visible and near-infrared ranges and a high albedo, has a significantly higher bulk density of 3.3 g/cm$^3$ (Descamps et al. 2008). Laver et al. (2009) reported that both components' spectra are similar within a 5-10% confidence interval and concluded that this binary system formed from the same material at the same time.

The spectroscopic study of a third binary system, (379) Huenna, a 98-km diameter primary with a 9-km satellite, which is also a C-type asteroid and member of the Themis collisional family, was recently presented by DeMeo et al. (2010) using observations taken under good viewing conditions with IRTF/SpeX. The near-infrared spectra of both components are identical, suggesting that the secondary is either a fragment of the primary or a fragment from the collision that created the Themis family.

At present, no spectroscopic studies of multiple asteroid systems reveal differences among system components. Shall we conclude that all these systems formed from the same material? It is too early to reach this conclusion, as the sample of multiple asteroids that we have studied spectroscopically is very small compared to the 195 known multiple asteroid systems. Our list of potential targets for component-resolved, near-infrared spectroscopy is also biased by the limits of adaptive optics systems, which can only resolve binary asteroids with an angular separation of at least 0.050 mas (for Keck II/OSIRIS) and a typical brightness ratio of 4-5 magnitude. Smaller or more distant binary asteroids, with closer components, cannot be studied using an integral field unit instrument on an 8-10m class telescope. One solution toward overcoming this limitation is to accurately photometrically characterize a binary asteroid system and to schedule photometric and spectroscopic observations when the components are in occultation configurations. Today, however, the orbits of a handful number of multiple asteroid systems are known well enough to predict their components' positions (Marchis et al. 2008, Marchis et al. 2010). To expand this study to a large number of binary asteroid systems, a significant effort in photometric surveying is still necessary. Whole-sky surveys like PS1 and the planned Pan-Starrs or LSST should soon provide the needed data. From the VOBAD database (Marchis et al. 2006), we estimate that such a study could be performed on up to 70 binary asteroids whose components have a size ratio of 1/5 or higher, the typical limit of detection by photometric surveys (Pravec et al. 2006).

Polishook et al. (2009) showed that, in the case of (90) Antiope, such photometric and spectroscopic studies could be conducted using 1m-class telescopes. We recommend pursuing such studies in the future, broadening them to longer wavelengths in near- and mid-infrared. It could be of particular interest to record high SNR, near-infrared spectra of (90) Antiope during occultation events, since Descamps et al. (2009) reported the presence of a large bowl-shape crater on one of the components. This crater may give us the opportunity to detect fresher material, not yet weathered by space, containing valuable information on the interior of (90) Antiope. Unfortunately, from our current orbital model, we found out that there will be only 2 epochs, centered on November 2018 and March 2025, offering the opportunities to observe (90) Antiope in mutual event configurations from Earth until 2032.

The present work added one more piece to the puzzle of (90) Antiope's origin, confirming that the two components have likely formed from the same body at the same time. The details of this formation—whether it involved tidal fission, splitting after an oblique impact, or any other scenarios not yet conceived—are unknown. (90) Antiope is the largest same-sized binary system currently known. As a member of the Themis collisional family, it is likely the oldest and largest rubble-pile binary asteroid in the main belt. Its study provides an opportunity to learn about the early solar system, when catastrophic disruptions in the main asteroid belt could have been frequent. Its uniqueness may be linked to a complex collisional history, as it could represent the aftermath of disruptions of larger rubble-pile asteroids. Numerical simulations, or most advantageously, in-situ characterizations from a spacecraft, could be the only way to fully understand the origin of this intriguing binary system and complete the (90) Antiope puzzle.

**Acknowledgement**

FM work was supported by the National Science Foundation under award number AAG-0807468. This research has made use of NASA's Astrophysics Data System and Mendeley$^{TM}$ software. This publication makes use of data products from the Two Micron All Sky Survey, which is a joint project of the University of Massachusetts and the Infrared Processing and Analysis Center/California Institute of Technology, funded by the National Aeronautics and Space Administration and the National Science Foundation. We are grateful to Todd Bradley and Keaton Burns for providing edits and comments that improved this manuscript and for C. Chapman and an anonymous reviewer for their valuable comments which improved significantly this manuscript.

**References**

Campins, H., Hargrove, K., Pinilla-Alonso, N., Howell, E. S., Kelley, M. S., Licandro, J., et al. (2010). Water ice and organics on the surface of the asteroid 24 Themis. *Nature*, *464*(7293), 1320-1321. doi: 10.1038/nature09029.

Castillo-Rogez, J. C., & Schmidt, B. E. (2010). Geophysical evolution of the Themis family parent body. *Geophysical Research Letters*, *37*(10), L10202. American Geophysical Union. doi: 10.1029/2009GL042353.

DeMeo, F. E., Binzel, R. P., Slivan, S. M., & Bus, S. J. (2009). An extension of the Bus asteroid taxonomy into the near-infrared. *Icarus*, *202*(1), 160-180. doi: 10.1016/j.icarus.2009.02.005.

DeMeo, F. E., Carry, B., Marchis, F. et al. (2010). (379) Huenna's Satellite: A Chip Off The Block. *American Astronomical Society*, *42*.

Descamps, P., Marchis, F., Michalowski, T., Vachier, F., Colas, F., Berthier, J., et al. (2007). Figure of the double Asteroid 90 Antiope from adaptive optics and lightcurve observations. *Icarus*, *187*(2), 482-499. doi: 10.1016/j.icarus.2006.10.030.

Descamps, P., Marchis, F., Pollock, J., Berthier, J., Vachier, F., Birlan, M., et al. (2008). New determination of the size and bulk density of the binary Asteroid 22 Kalliope from observations of mutual eclipses. *Icarus*, *196*(2), 578-600. doi: 10.1016/j.icarus.2008.03.014.

Descamps, P., Marchis, F., Michalowski, T., Berthier, J., Pollock, J., Wiggins, P., et al. (2009). A giant crater on 90 Antiope? *Icarus*, *203*(1), 102-111. doi: 10.1016/j.icarus.2009.04.022.

Emery, J. P. Burr, D. M., Cruikshank, D. P., Brown, R. H., Dalton, J. B. (2005). Near-infrared (0.8-4.0 µm) spectroscopy of Mimas, Enceladus, Tethys, and Rhea. *Astronomy and Astrophysics* 435, 1, 353-362 doi:10.1051/0004-6361:20042482.

Florczak, M., Lazzaro, D., Mothe-Diniz, T., Angeli, C. A., & Betzler, A. S. (1999). A spectroscopic study of the Themis family. *Astronomy and Astrophysics Supplement Series*, *134*(3), 463-471. doi: 10.1051/aas:1999150.

Eisenhauer, F. (2003). *SINFONI - Integral field spectroscopy at 50 milli-arcsecond resolution with the ESO VLT*. Proceedings of SPIE (Vol. 4841, pp. 1548-1561). SPIE. doi: 10.1117/12.459468.

Hansen, A. T., Arentoft, T., & Lang, K. (1997). The Rotational Period of 90 Antiope. *Minor Planet Bulletin*, *24*, 17.

Marchis, F., Descamps, P., Baek, M., Harris, A., Kaasalainen, M., Berthier, J., et al. (2008). Main belt binary asteroidal systems with circular mutual orbits✶. *Icarus*, *196*(1), 97-118. doi: 10.1016/j.icarus.2008.03.007.

Marchis, F., Baek, M., Berthier, J., Descamps, P., Hestroffer, D., Kaasalainen, M., Vachier, F., 2006c. Large adaptive optics of asteroids (LAOSA): Size, shape, and occasionally density via multiplicity. In: Workshop on Spacecraft Reconnaissance of Asteroid and Comet Interiors. Abstract 3042.

Marchis, F. et al. (2010). A Dynamical Solution of the Triple Asteroid System (45) Eugenia. Icarus, Volume 210, Issue 2, p. 635-643

Marzari, F. (1995). Collisional Evolution of Asteroid Families. *Icarus*, *113*(1), 168-187. doi: 10.1006/icar.1995.1014.

Merline, W. J. et al. (2000). Discovery of Companions to Asteroids 762 Pulcova and 90 Antiope by Direct Imaging. *American Astronomical Society*, *32*.

Murchie, S.L. and Erard, S. 1996. Spectral Properties and Heterogeneity of PHOBOS from Measurements by PHOBOS 2. *Icarus* 123, no. 1: 63-86. doi:10.1006/icar.1996.0142.

Murchie, S. L. et al. 2008. MRO/CRISM Observations of Phobos and Deimos. *39th Lunar and Planetary Science Conference*. http://adsabs.harvard.edu/abs/2008LPI....39.1434M.

Nesvorný, D., Bottke, W. F., Vokrouhlický, D., Morbidelli, A., & Jedicke, R. (2005). Asteroid families. *Proceedings of the International Astronomical Union*, *1*(S229), 289-299. Retrieved from http://journals.cambridge.org/abstract_S1743921305006800.

Pang, K. D., Pollack, J. B. , Veverka, J. , Lane A. L., and Ajello, J. M.. 1978. The Composition of Phobos: Evidence for Carbonaceous Chondrite Surface from Spectral Analysis. *ScienceNew Series* 199, no. 4324: 64 - 66. http://www.jstor.org/stable/1745507.

Polishook, D., Brosch, N., Prialnik, D., & Kaspi, S. (2009). Simultaneous spectroscopic and photometric observations of binary asteroids. *Meteoritics & Planetary Science*, *44*(12). Retrieved from http://adsabs.harvard.edu/abs/2009M&PS...44.1955P.

Pravec, P., Scheirich, P., Ku, P., Mottola, S., Hahn, G., Brown, P., et al. (2006). Photometric survey of binary near-Earth asteroids. *Icarus*, *181*, 63-93. doi: 10.1016/j.icarus.2005.10.014.

Rivkin, A. S., & Emery, J. P. (2010). Detection of ice and organics on an asteroidal surface. *Nature*, *464*(7293), 1322-3. doi: 10.1038/nature09028.

**Figure 1:** Schematic view of several Type-2 ("similarly-sized" or double) binary asteroid systems known in the main belt, extracted from VOBAD (Marchis et al. 2006). Due to the large size (D~90 km) of its components, (90) Antiope is obviously a unique double binary and the only one which can be resolved using current 8-10m class telescopes with adaptive optics systems.

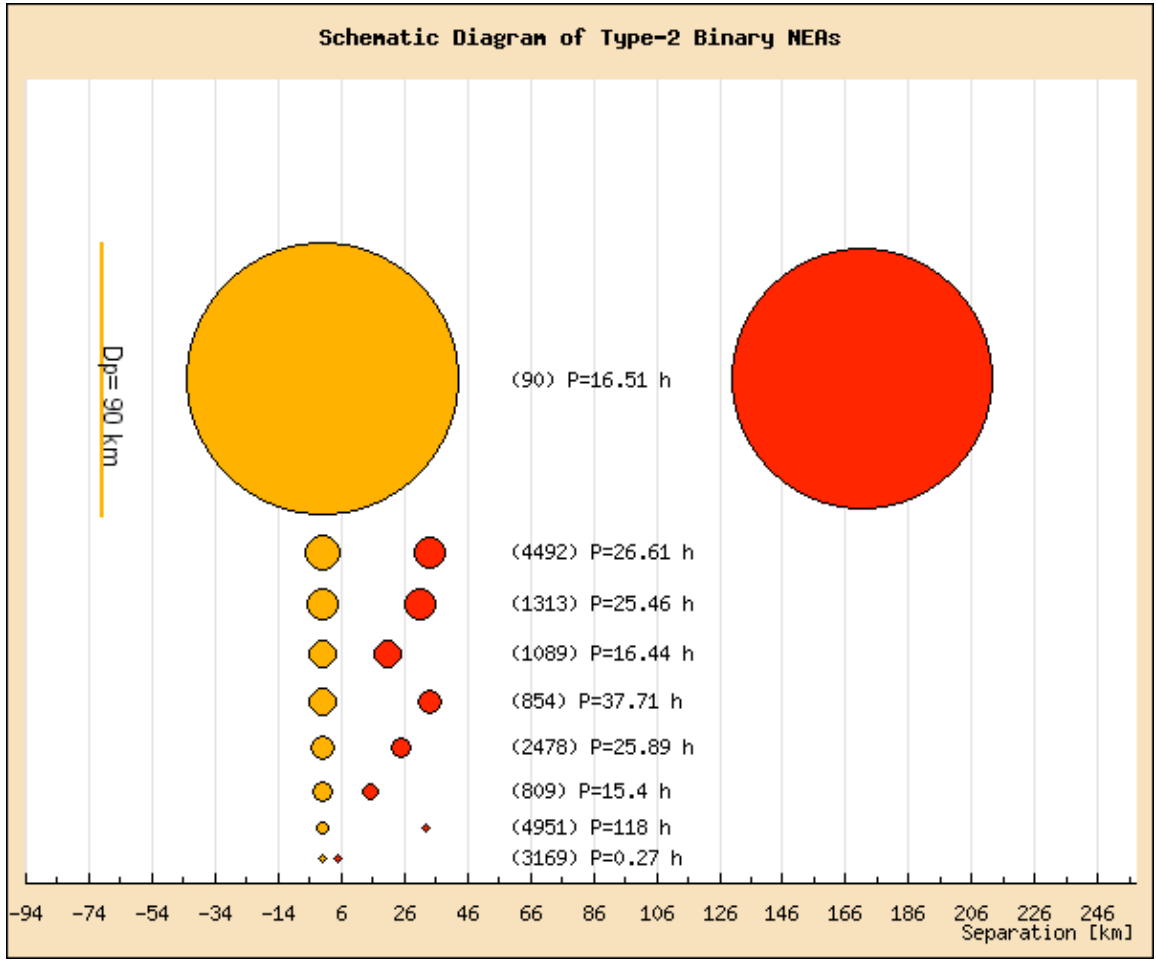

**Figure 2:** Typical visible and near-infrared spectra of several C-complex, X-type, and S-type asteroids from DeMeo et al. (2009), normalized at 1 µm. The broader wavelength range and the presence of absorption bands in the near-infrared (λ>1 µm) should ease the identification of spectral differences linked to surface composition for the two components of (90) Antiope.

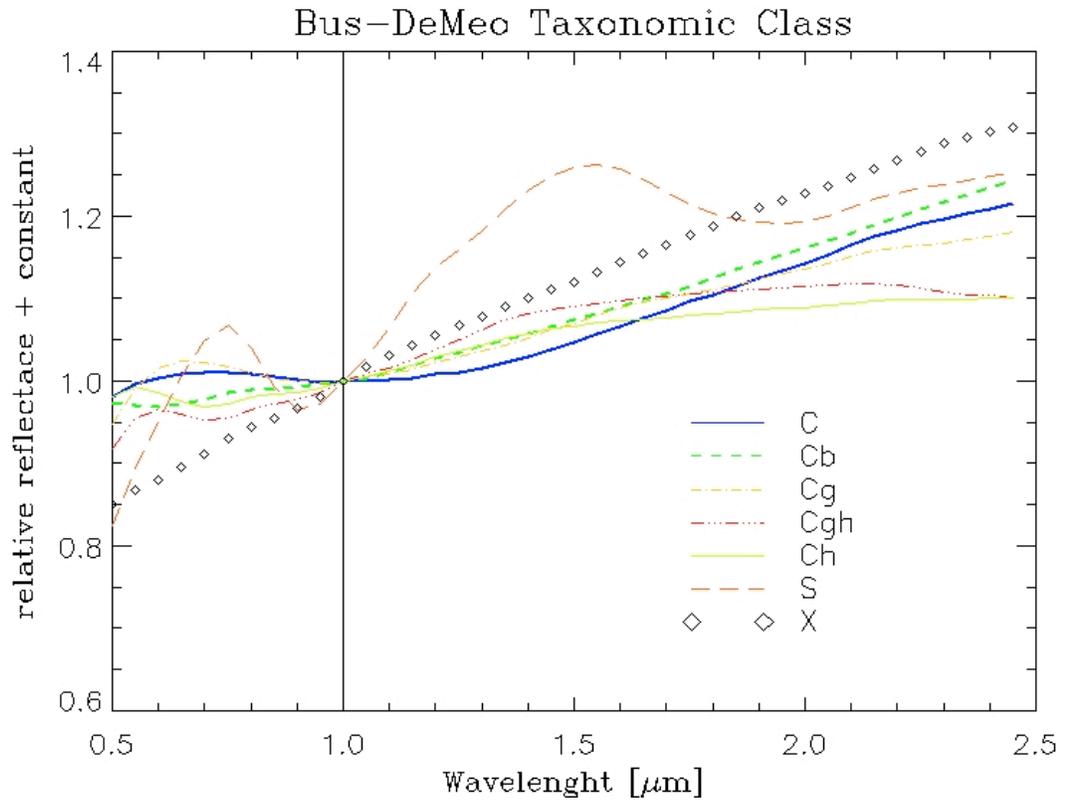

**Figure 3:** Slices of the data cube extracted from the H+K grating (at 1.55 µm) and the J grating (at 1.13 µm) showing the reconstructed images for 3 epochs of observations in the first two columns. Both components of (90) Antiope are visible in the Feb 21 2009 data. These two frames were recorded with a time difference of ~1.25h, explaining the different positions of the components. The last column shows the predicted positions from the Descamps et al. (2009) model, which is in excellent agreement with the H+K observations. North is up and East is left on these images. Feb 03 data in both grating modes and Feb 01 data in J mode are of poor quality and were not included in our data analysis.

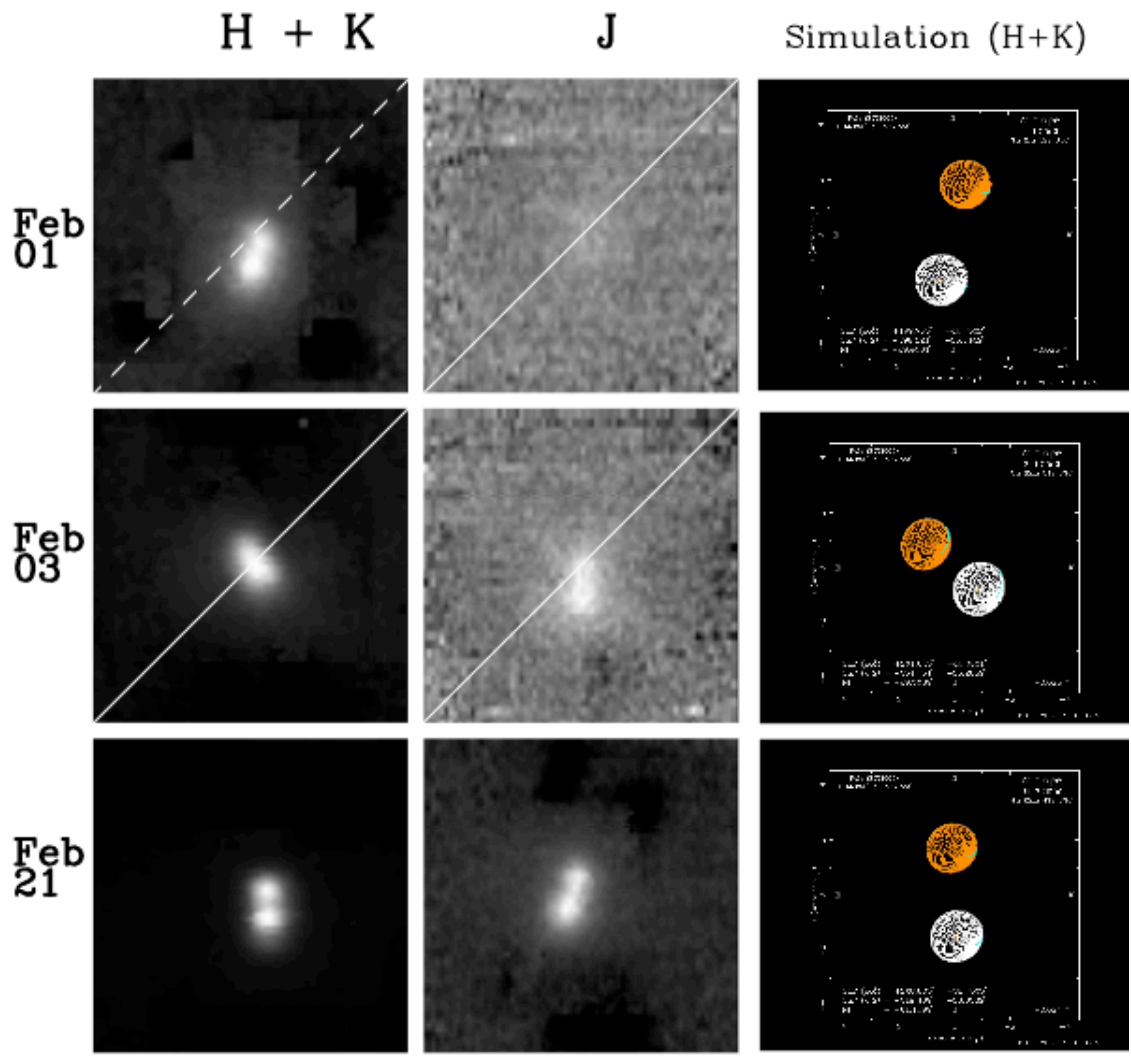

**Figure 4:** 0.89 × 0.89 arcsec images extracted from the H+K grating cube of data taken on Feb 21 (top) and Feb. 01 (bottom) at various wavelengths from 1.4475 to 2.107 µm and displayed with the same scale. Feb 21 data have a better SNR due to better seeing conditions. This figure illustrates the variability in quality (angular resolution and SNR) into one cube of data and between nights.

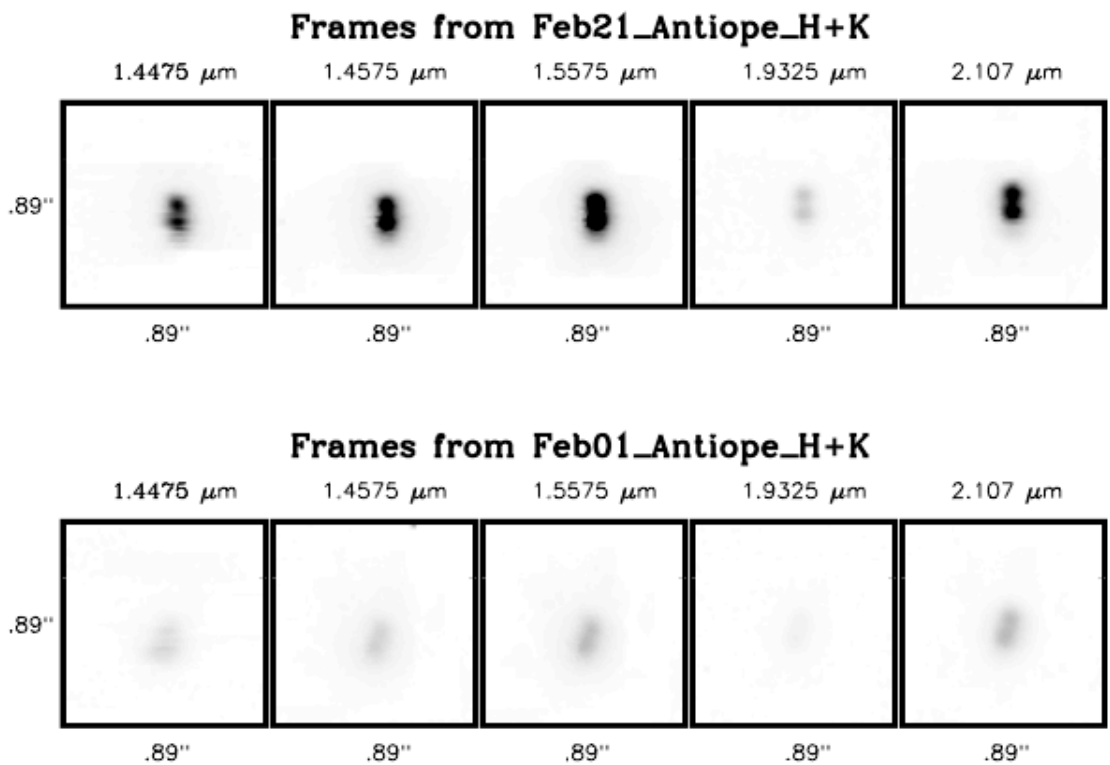

**Figure 5:** SNR, radial and tangential FWHMs calculated for two epochs of observations (Feb 21 and Feb 01) in J and H+K bands for both components of (90) Antiope. These plots illustrate the variability of the data in terms of quality (sensitivity and angular resolution) along one mode of observations and between these two nights. Feb 01 data in H+K and Feb 21 data in J band have similar, low SNRs (<50) and similar FWHMs (~6 pixels) profiles, implying that they are of similar quality.

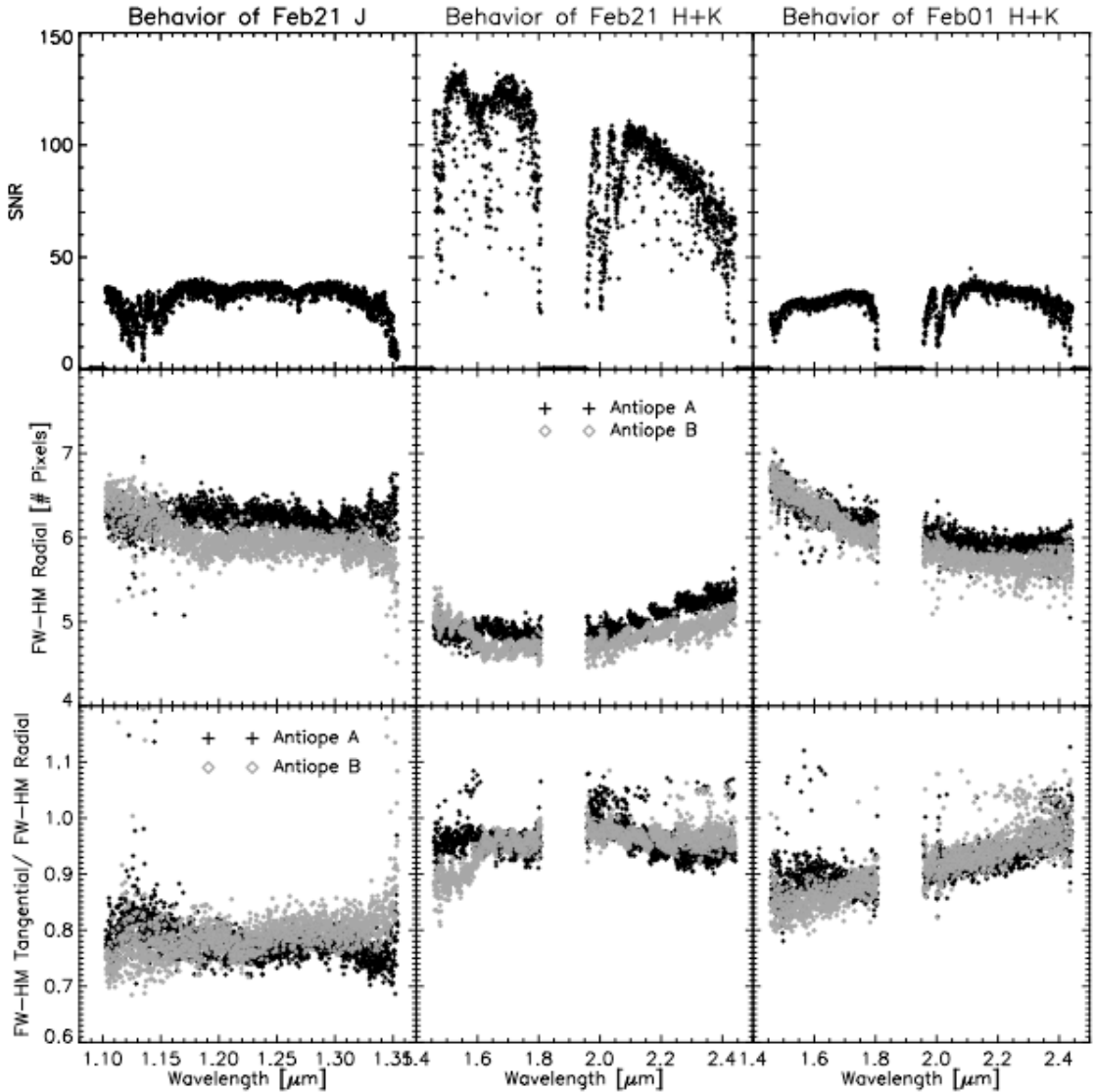

**Figure 6:** An example of the fitting process applied on a J-band frame (1.22 µm) taken on Feb 21. The inset images show the extracted slide, the fit with our Gaussian functions (a wide and a narrow Gaussian functions per component, see Section 3.1) and the residuals, from left to right. A cut along the radial direction is plotted on the top figure. The Gaussian fit functions reproduce the profile of the components quite well. The residuals in the central part of the image are less than 3% of the total flux value.

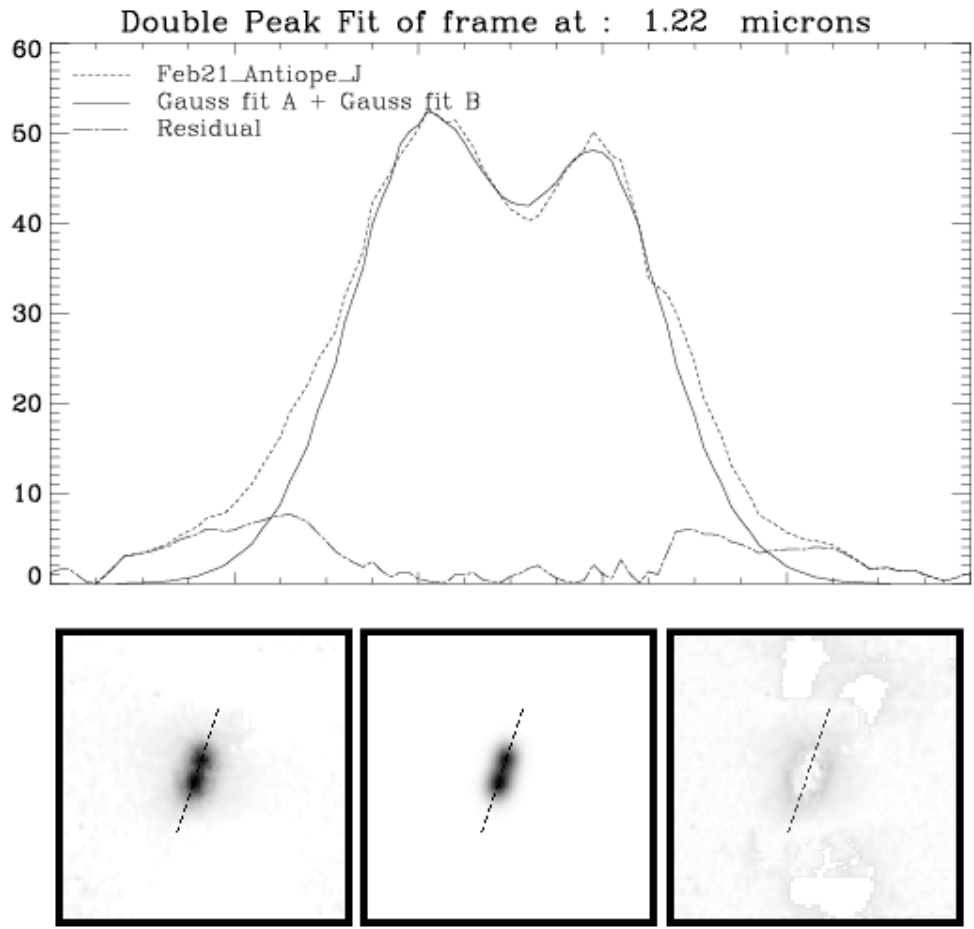

**Figure 7:** Extracted spectra of each component of (90) Antiope, rebinned to a low spectral resolution of 150 in H+K grating and 200 in J grating and shifted by an arbitrary value of ± 0.2. Antiope A corresponds to the larger component and Antiope B to the smaller component. The average spectrum from both components' spectra is plotted in the middle. For comparison we plotted an integrated spectrum recorded with IRTF/Spex on July 22 2006. The Feb 21 spectrum is identical in slope to the integrated spectrum, validating our extraction algorithm. Feb 01 spectra are significantly redder than the IRTF/Spex spectrum, suggesting that the fitting algorithm is unable to properly extract the flux due to the poor quality of the data taken in low seeing conditions.

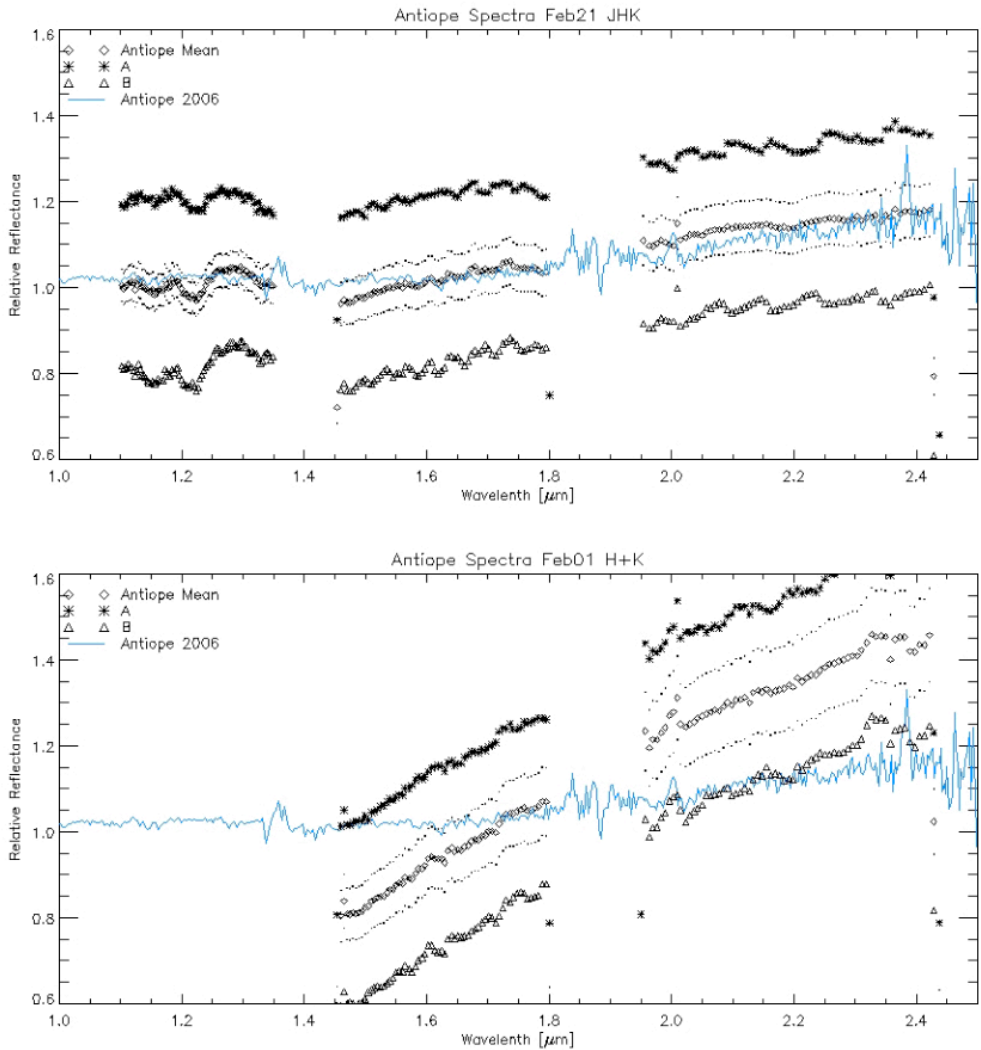

**Figure 8:** Example of synthetic images with SNRs of 25 and 60 and *proximities* from 0.8 to 1.2 created for our Monte-Carlo simulation. To mimic the real observations of (90) Antiope as closely as possible, these 2 components have a flux ratio of 0.90 and are depicted at 60 degrees counter clockwise. When the *proximity*, which is equal to the angular separation of the components divided by the FWHM of the components, is greater than or equal to 1, the binary system is fully resolved.

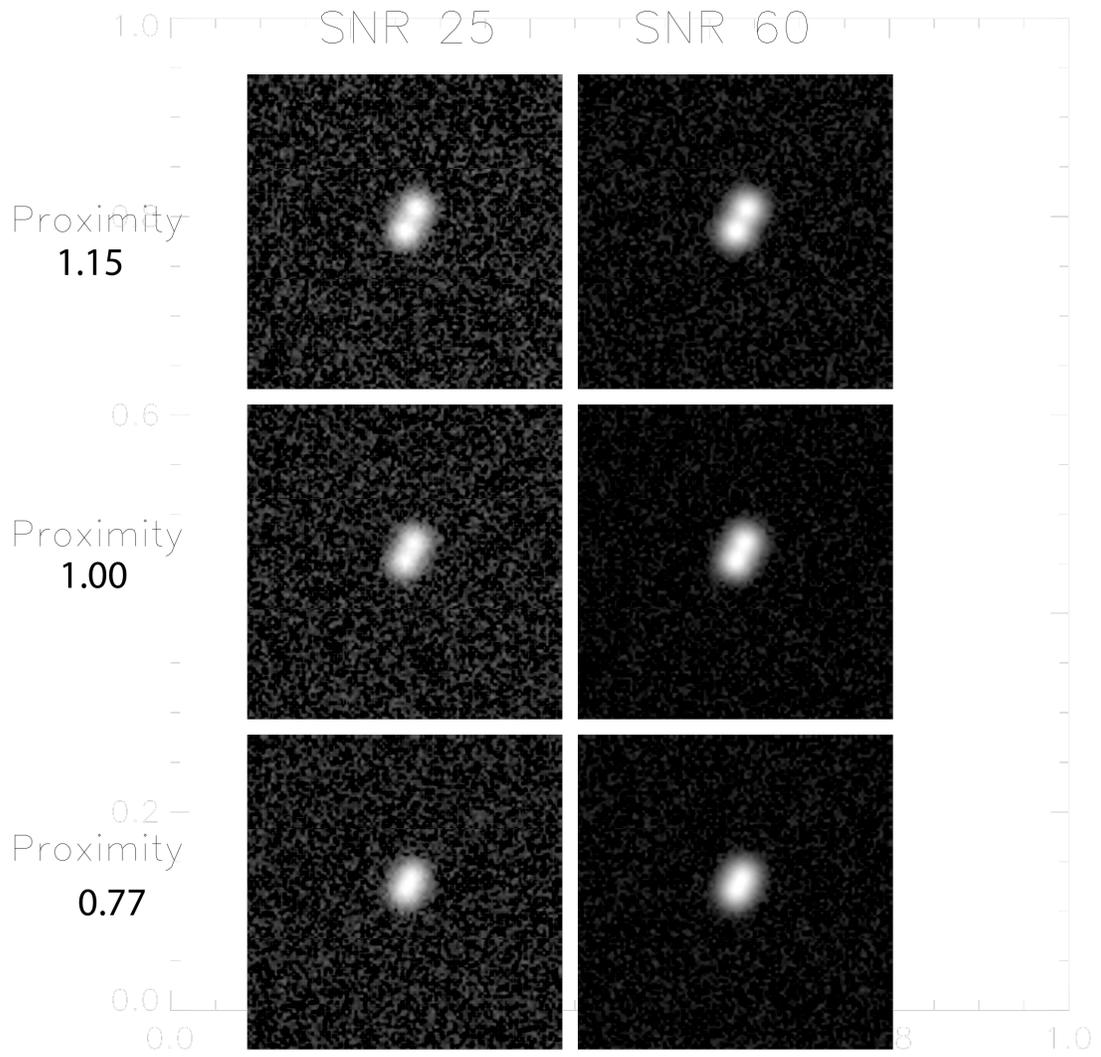

**Figure 9:** Accuracy estimates on the flux ratio derived from our Monte-Carlo analysis of artificial images. The accuracy varies from 3% to 9% for data with a SNR from 25 to 160 and a *proximity* from 0.8 to 1.2.

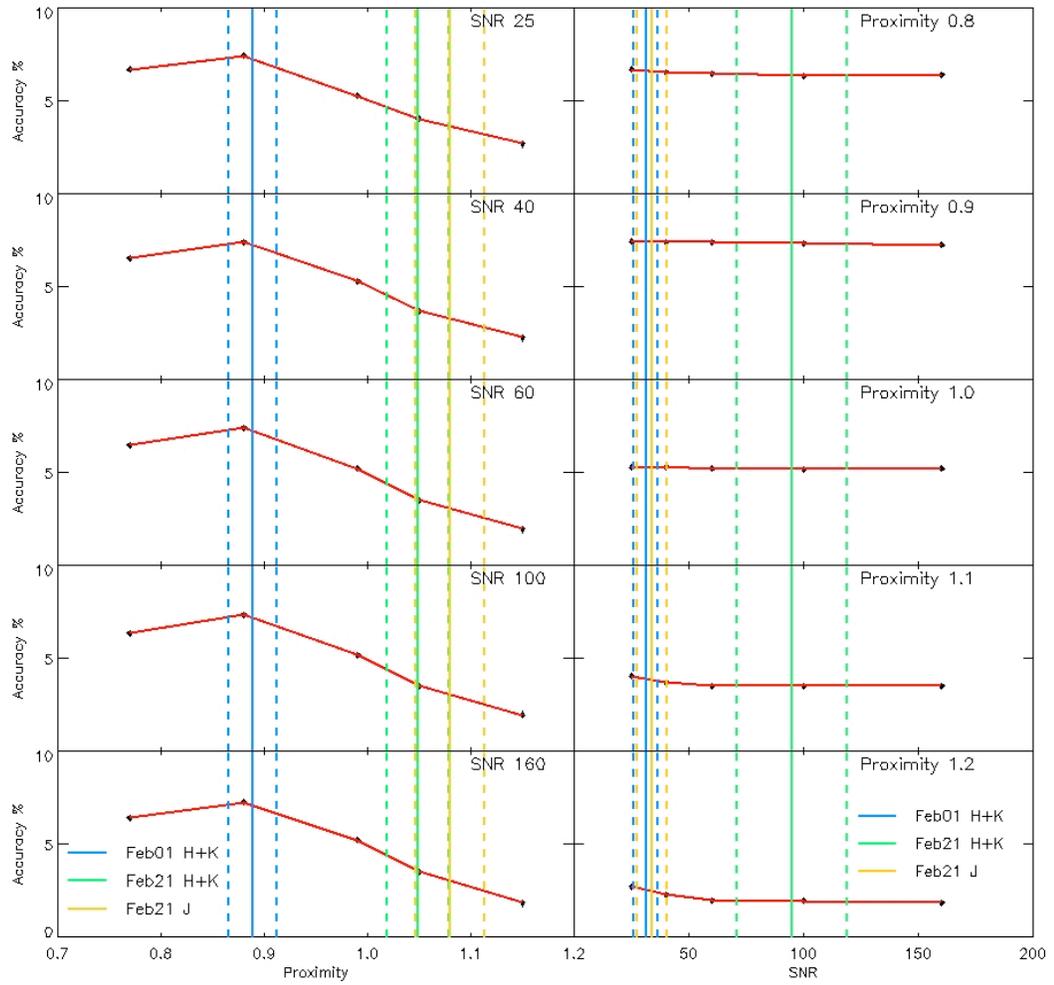

**Figure 10:** Final 1.1 to 2.4 μm spectra, with formal error bars, of both components of (90) Antiope recorded on Feb 21, 2009. We overplot on these spectra the average spectrum of C-, Cb-, Cg- and Cgh- types asteroids from the de Meo et al. (2009) taxonomic classification and their 1-sigma variations. The Ch-type average spectrum, which is not shown here, does not fit our observations, Based on our extraction method, Antiope A & B seem to be C- or Cb- type asteroids.

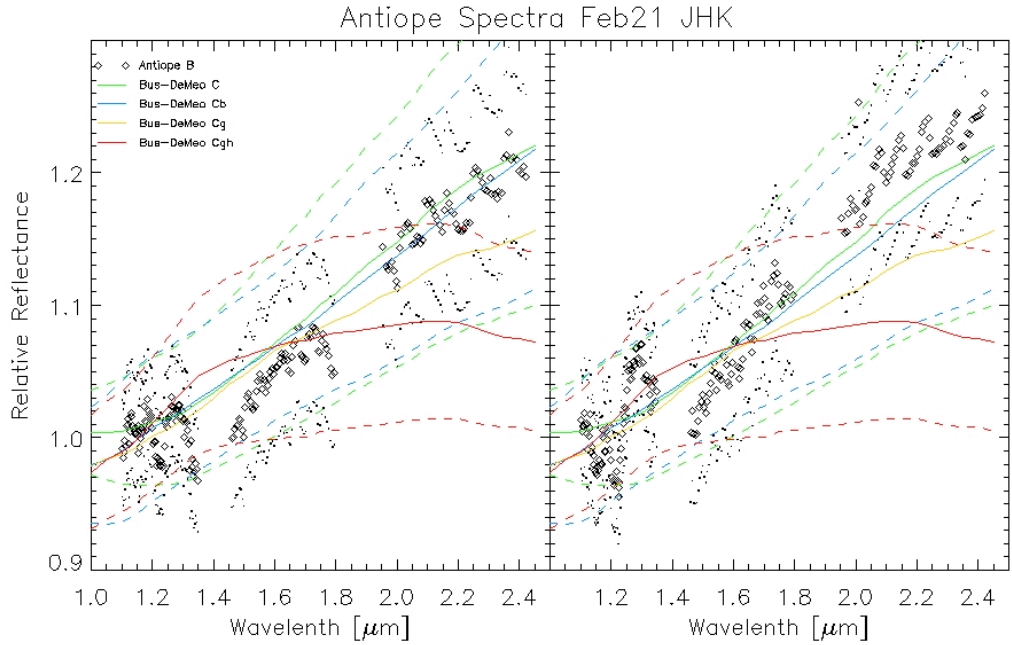

**Figure 11:** Spectral flux ratio between Antiope B and Antiope A derived from our photometric extraction and error analysis. The ratio is remarkably constant, with a best fit of 0.87 and a 1-sigma residual error of 0.07, implying that both components have an identical reflectance spectrum in this near-infrared wavelength range.

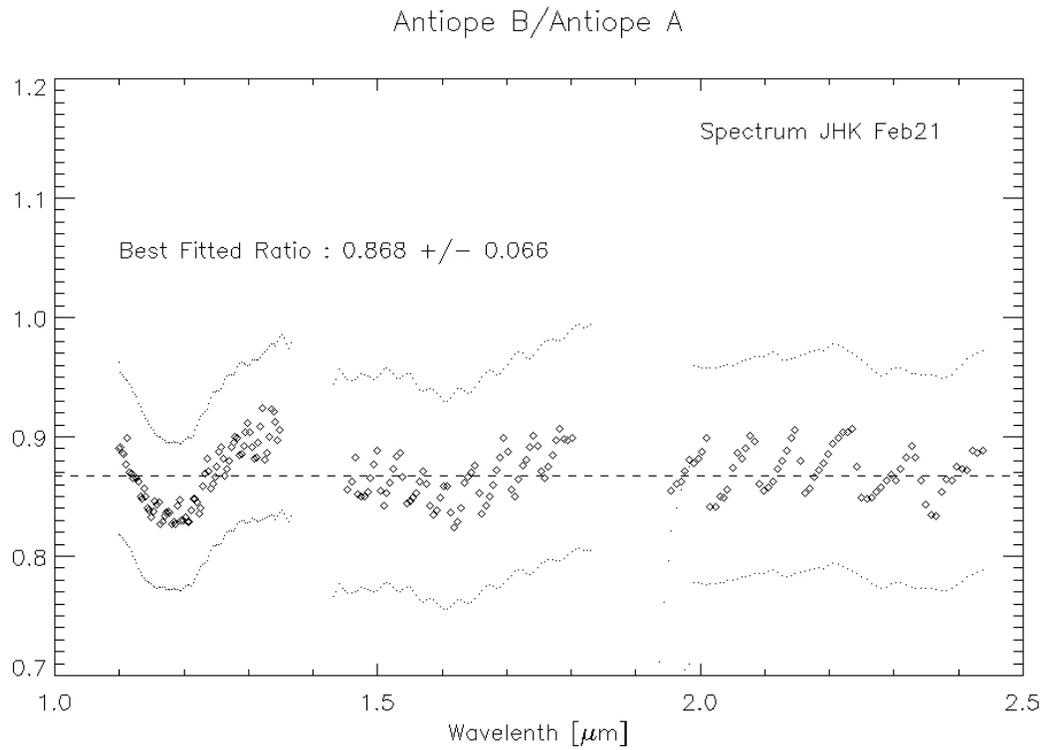

**Table 1:** Observational logs of the VLT SPIFFI/OSIRIS observations recorded in Feb 2009. Only one observation, taken on Feb 21, was recorded under the criteria of atmospheric quality and predicted angular separation indicated in our proposal.

| Target | Date Time in UT | Exp. time in s | Grating | Airmass | Seeing in arcsec | Note |
|---|---|---|---|---|---|---|
| Antiope | 2/1/09 7:00 | 300 | H+K | 1.203 | 1.25 | |
| Antiope | 2/1/09 5:42 | 300 | J | 1.273 | 1.24 | Poor seeing quality below |
| BD+11-2282 | 2/1/09 6:44 | 60 | H+K | 1.22 | 1.19 | |
| BD+11-2282 | 2/1/09 8:01 | 60 | H+K | 1.301 | 0.98 | |
| BD+11-2282 | 2/1/09 6:28 | 60 | J | 1.223 | 1.26 | requested |
| BD+11-2282 | 2/1/09 5:19 | 60 | J | 1.318 | 1.32 | |
| Antiope | 2/3/09 6:16 | 300 | H+K | 1.217 | 0.75 | |
| Antiope | 2/3/09 7:37 | 300 | J | 1.237 | 0.78 | Observations not requested, |
| BD+11-2282 | 2/3/09 7:01 | 60 | H+K | 1.228 | 1.37 | |
| BD+11-2282 | 2/3/09 5:23 | 60 | H+K | 1.292 | unknown | |
| BD+11-2282 | 2/3/09 7:11 | 60 | J | 1.236 | 1.18 | components too close. |
| BD+11-2282 | 2/3/09 8:22 | 60 | J | 1.379 | 1.04 | |
| Antiope | 2/21/09 4:29 | 300 | H+K | 1.264 | 0.56 | |
| Antiope | 2/21/09 5:49 | 300 | J | 1.231 | 0.61 | Optimal angular separation, |
| BD+11-2282 | 2/21/09 5:16 | 60 | H+K | 1.221 | 0.70 | |
| BD+11-2282 | 2/21/09 4:07 | 60 | H+K | 1.301 | 0.73 | |
| BD+11-2282 | 2/21/09 5:27 | 60 | J | 1.22 | 0.64 | seeing. |
| BD+11-2282 | 2/21/09 6:32 | 60 | J | 1.28 | 0.67 | |